\documentstyle[rotate,psfig,twocolumn]{mn}
\title[3C254: a highly asymmetric quasar]{3C254: MERLIN observations of a highly asymmetric quasar}

\author[P. Thomasson et al.]{P. Thomasson$^{1}$, D.J. Saikia$^{1,2}$ and T.W.B. Muxlow$^{1}$ \\
$^{1}$ The University of Manchester, Jodrell Bank Observatory, Macclesfield, Cheshire, SK11 9DL \\
$^{2}$ Tata Institute of Fundamental Research, National Centre for Radio
Astrophysics, Ganeshkhind, Pune 411 007, India \\
}

\date{Received:}
\begin{document}
\maketitle

\begin{abstract}
Multifrequency, high-resolution radio observations of the quasar 3C254 using MERLIN 
are presented. The quasar has a highly asymmetric radio structure,
with the eastern component of the double-lobed structure being much closer to the nucleus
and significantly less polarized than the western one. However, the two lobes are more 
symmetric in their
total flux densities. The observations show the detailed structure of the hotspots which 
are very different on opposite sides of the radio core, reveal no
radio jet and suggest that the oppositely-directed jets may be intrinsically asymmetric.  
\end{abstract}


\section{Introduction}
The majority of high-luminosity extragalactic radio sources selected at
a low frequency tend to be symmetric in the brightness and location of the 
components relative to the nucleus of the parent galaxy. 
However, a small but significant fraction are highly 
asymmetric.  The asymmetries of the lobes of emission are 
important because they could provide useful insights into the environments
of sources (e.g. Pedelty et al. 1989; McCarthy, van Breugel \& Kapahi 1991), 
the interaction of jets with external clouds or galaxies, and also
provide tests of the orientation-based unified scheme for radio galaxies 
and quasars (Barthel 1989). For example, the structure of the highly-asymmetric double-lobed
source B0500+630 appears to be inconsistent with the unified scheme and suggests that 
there is an intrinsic asymmetry in the oppositely-directed jets from the nucleus
(Saikia et al. 1996). The most extreme form of asymmetry is when the source is
completely one-sided with radio emission on only one side of the nucleus. 
Although most of these objects are core-dominated and their apparent asymmetry is
likely to be due to bulk relativisitic motions in the extended lobes of emission,
there does appear to be a number of weak-cored one-sided sources which are difficult
to reconcile with the simple relativistic beaming scenario (Saikia et al. 1990). 

As part of a study of highly asymmetric radio sources, the quasar 
3C254 has been observed at radio wavelengths with high resolution using MERLIN. 
The results of these observations are presented in this paper. 3C254 is identified 
with a quasar at a redshift of 0.73612$\pm$0.00160 (NASA Extragalactic Database)
so that 1 arcsec corresponds to 7.290 kpc 
(H$_o$=71 km s$^{-1}$ Mpc$^{-1}$, $\Omega_m$=0.27, $\Omega_\Lambda$=0.73,
Spergel et al. 2003).  

The radio source has an overall angular extent of 13.5 arcsec (98 kpc)
with the eastern component being much closer to the nucleus. The ratio of the
separations of the outer hotspots, r$_D$, defined to be $>$1, is 7.1 while
the flux density ratio of the corresponding lobes from the MERLIN {\it L}-band images presented
here is $\sim$0.55. Earlier Very Large Array (VLA) and MERLIN observations with lower
resolution have shown the highly asymmetric structure and a bridge of
emission connecting the two lobes (e.g. Owen \& Puschell 1984; Liu \& Pooley 1991;
Liu, Pooley \& Riley 1992; Reid et al. 1995).  The observations by Liu \& Pooley have also shown 
that the eastern lobe is more depolarized, and has a steeper spectrum, consistent 
with the relationship reported by them that the more depolarized lobe tends to have 
a steeper radio spectrum. This relationship is
significantly stronger for smaller sources, but is similar for both radio galaxies and quasars
suggesting that, in addition to Doppler effects, there are intrinsic differences between the
lobes on opposite sides (cf. Ishwara-Chandra et al. 2001).  The large separation ratio,
the distortion in the eastern lobe and the stronger depolarization all 
suggest that the jet on the eastern side is colliding with a dense gas cloud
which is influencing the properties and appearance of the radio source. 

Evidence of the gas clouds has been reported by Forbes et al. (1990)  who found
this steep-spectrum quasar to be embedded in a spectacular, extended, emission-line 
region emitting strongly in [O{\sc ii}] $\lambda$3727 and 
[O{\sc iii}] $\lambda$5007. Forbes et al. have also suggested that the emitting gas
is at high pressure, consistent with confinement by a hot intracluster medium.
Subsequent observations have shown that the optical continuum and line emission are
extended along the radio axis (Bremer 1997; Crawford \& Vanderriest 1997),
illustrating the so-called alignment effect which is seen in powerful radio galaxies.
Bremer also shows that part of the extended emission line region which is close to the 
eastern lobe has a blueshift of $\sim$650 km s$^{-1}$ relative to the nuclear emission 
along a position angle of 105$^\circ$. There is also an overdensity of galaxies within 
approximately 10 arcsec of the quasar, suggesting that 3C254 lies at the centre of a compact
group of galaxies. High-resolution X-ray observations with the Chandra Observatory
show X-ray emission from the quasar and the western hotspot, and demonstrate that
the source does not lie in a hot, massive cluster (Donahue, Daly \& Horner 2003).

\begin{figure*}
\begin{center}
\vbox{
\hspace{0.0cm}
\psfig{file=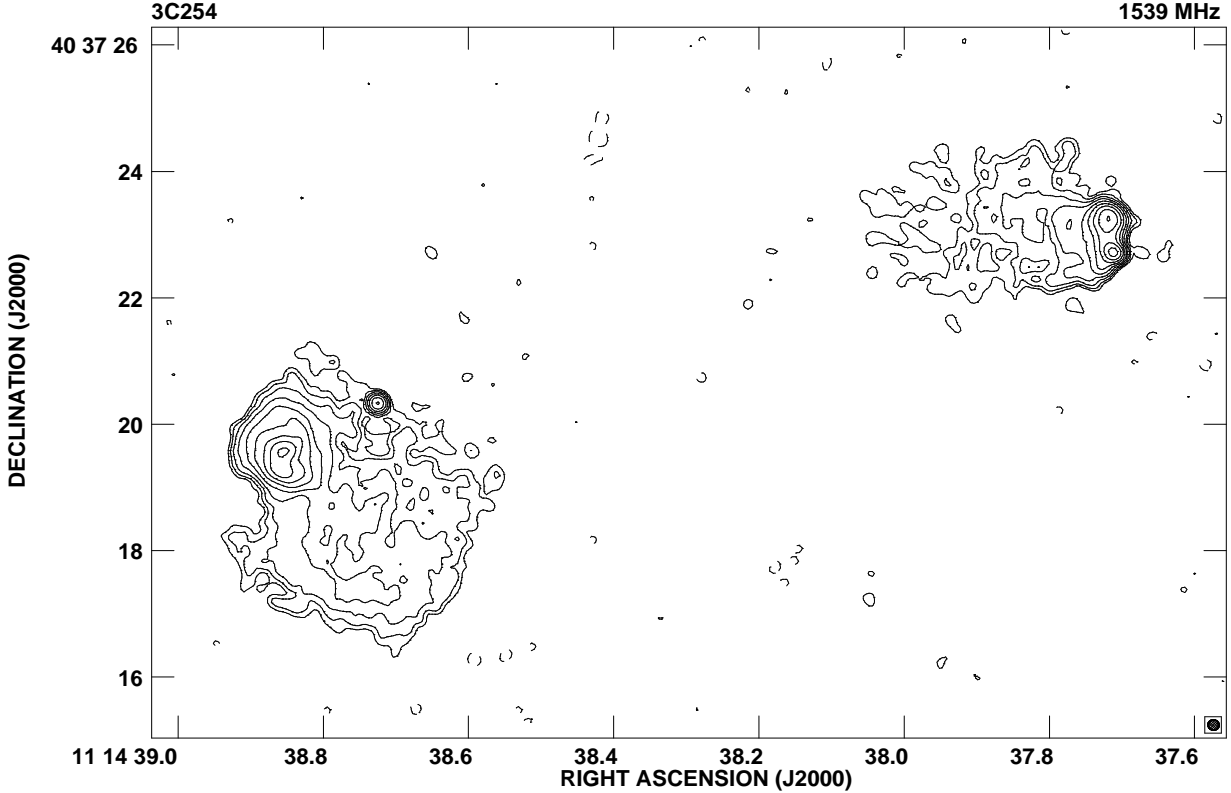,width=6.5in,angle=-90}
}
\end{center}
\caption{The MERLIN images of 3C254 at 1539 MHz with an angular resolution of $\sim$0.17 arcsec.
Peak brightness: 142 mJy/beam; Contours: 0.4$\times$($-$1, 1, 2, 4, 8, 16 $\ldots$) mJy/beam.
}
\end{figure*}

In this paper, high-resolution MERLIN observations at {\it L}- and {\it C}-bands
are presented. These observations were made to examine the structure and asymmetry 
of the hotspots on opposite sides and determine any possible intrinsic asymmetry 
in the oppositely-directed jets.
The observations and analyses of the data are described in Section 2, and the 
results are presented in Section 3. Possible explanations for the structure are
discussed in Section 4, while the conclusions are summarised in Section 5.

\section{Observations and analyses}

The MERLIN observations described in this section have been made at {\it L}- and 
{\it C}-band frequencies. The observations have been made at different 
frequencies within each band, and at {\it C}-band the data have also been combined 
with VLA data to improve the {\it uv}-coverage and thus the resulting images. The flux calibration 
source for the observations within both bands was 3C286, its flux densities at the differing 
frequencies being calculated from the values given in the VLA calibrator source catalogue, which 
is based on the work of Baars et al. (1977).   The flux densities of the point source baseline 
calibrators (DA193 for the {\it L}-band observations; OQ208 for the {\it C}-band observations) have been 
determined from a comparison of their MERLIN short-spacing amplitudes with those of 3C286.   
3C286 is assumed to have a position angle for its linear polarization of 32.4 degrees.   
The phase reference source used for all the observations was B1115+416.   
Its J2000 position, obtained from the 
Interferometer Phase Calibration Source list I of Patnaik et al. (1992), 
is 11:17:53.33390 +41:20:16.2761.  The processing of the data
was carried out using Jodrell Bank data editing and calibration software and the MERLIN pipeline, 
the latter using Astronomical Image Processing Software {\tt AIPS} tasks. 

\subsection{{\it L}-band observations}
As indicated above, in order to improve the {\it uv} coverage and also to determine the polarization 
properties at two neighbouring {\it L}-band frequencies, 
3C254 was observed with MERLIN at 1420 MHz and 
1658 MHz on 3rd March 1996. Seven MERLIN telescopes were configured in the array, which included the 
Wardle telescope just prior to its decommissioning
and which therefore provided a minimum baseline length of 6 km in comparison with a minimum of 12 km 
at the present time. 
The observing frequency was switched between the two 
frequencies every 6 minutes and the observing cycle times between the target source (3C254) and 
its phase-reference (including telescope drive times) were in the ratio of 7.5:4.5. This resulted 
in $\sim$3.5 minutes on 3C254 at each frequency within the total 12 minute cycle time. 3C254 and its 
phase-reference were observed for a total of 17 hours in all four polarizations (LL, LR, RL and RR), 
resulting in $\sim$5 hours actually on source at each frequency.  
The flux densities of 3C286 which 
were used at 1658 MHz and 1420 MHz were 13.639 Jy and 14.733 Jy respectively.   
A final total-intensity image was produced at a mean `effective' 
frequency of 1539 MHz after scaling both the 1658 MHz and 1420 MHz data appropriately before combination. 
Polarisation images were produced at the two individual frequencies of 1420 MHz and 1658 MHz. 

\subsection{{\it C}-band observations}
3C254 was observed with MERLIN at the {\it C}-Band frequencies 4562 and 4874 MHz on 24, 25 
and 26 November 2002.   The total on-source integration times at the two frequencies were 
different and considerably less than a quarter of the total elapsed time as a result of system 
failures and the necessity to edit out data in which ionospherically induced, very high 
phase rates occurred. The observing frequency was switched every 10 minutes, with cycle times 
for observations of 3C254 and its phase reference at each frequency being in the ratio of 
6.5 to 3.5, giving $\sim$6 minutes on 3C254 for each ten minute period.  
As for the {\it L}-band observations, 3C254 and its phase-reference were observed in 
all four polarizations and the total time on-source 
at each frequency was $\sim$6 hours over the total observing period. Total-intensity and 
polarization images were produced at the same resolution for the two frequencies. 
The flux densities of 3C286 which were used at 4562 MHz and 4874 MHz were 7.807 Jy and 7.494 Jy 
respectively, although a correction had to be made to the data to take account of the fact that 
3C286 is $\sim$4\% resolved even on the shortest MERLIN baseline (12 km) at {\it C}-band frequencies.   
The data at the lower frequency were appropriately scaled to take account of the source spectral 
index and combined with that at the higher frequency to produce a combined image at an 
effective frequency of 4874 MHz.   
Finally, calibrated VLA data at a frequency of 4885 MHz, kindly provided 
by J. Riley and G. Pooley, were re-imaged and appropriately scaled and combined with the MERLIN 
data to produce a MERLIN+VLA image at 4874 MHz.

\section{Discussion}

Some of the observational parameters and observed values from the radio
observations are summarised in Table 1.

\subsection{The overall radio structure}

Earlier observations of 3C254 with an angular resolution of approximately 0.3$-$1 arcsec
show the two prominent lobes located very asymmetrically on opposite sides of the
quasar, a weak nucleus  and a cigar-shaped bridge of emission connecting the two
outer lobes (e.g. Owen \& Pushell 1984; Liu \& Pooley 1991; Reid et al. 1995). The
overall structure is reminescent of 3C459, a highly asymmetric radio galaxy with
a starburst which has been observed as part of this study of very asymmetric radio
sources (Thomasson, Saikia \& Muxlow 2003).

The MERLIN image of 3C254 at 1539 MHz with an angular resolution of 0.17 arcsec
(Fig. 1), shows the well-known double-lobed structure and the central component.
In addition, it reveals further details of the structure of the hotspots and the
lobes of extended emission in both the western and eastern lobes. The western lobe
consists of two hotspots, the primary and more compact one being towards the south.
The secondary hotspot and the diffuse emission from the lobe lies north of the
axis joining thc core to the primary hotspot. The peak of emission on the eastern
lobe appears to be less compact than the primary hotspot on the western lobe,
with the lobe emission being extended towards the south-west.

The western lobe is more strongly polarized with the peak in the hotspot being
$\sim$20\% polarized at 1420 MHz. There is no significant polarization from the peak
of emission in the eastern lobe, the 3$\sigma$ upper limit being $\sim$0.4\%, but there does appear to be polarized emission along the eastern rim of the diffuse `tail'.   However, the magnitude of this is very uncertain as it is only just above the noise level and requires confirmation from more sensitive observations (Fig. 2). 
Fig. 3 shows the MERLIN images of both the lobes of emission at 1658 MHz with an angular
resolution of 0.15 arcsec and with the polarization vectors
superimposed on them. As in the images at 1420 MHz, the hotspots in the western lobe
show significant polarization while there is no significant polarized signal from the peak of
emission in the eastern lobe, the upper limit being $\sim$0.6\%, although, once again, there is an indication of polarized emission along the eastern rim of the diffuse `tail'. The
degree of polarization in the peak of emission in the western lobe is $\sim$23\%.
                                                                                
\begin{figure*}
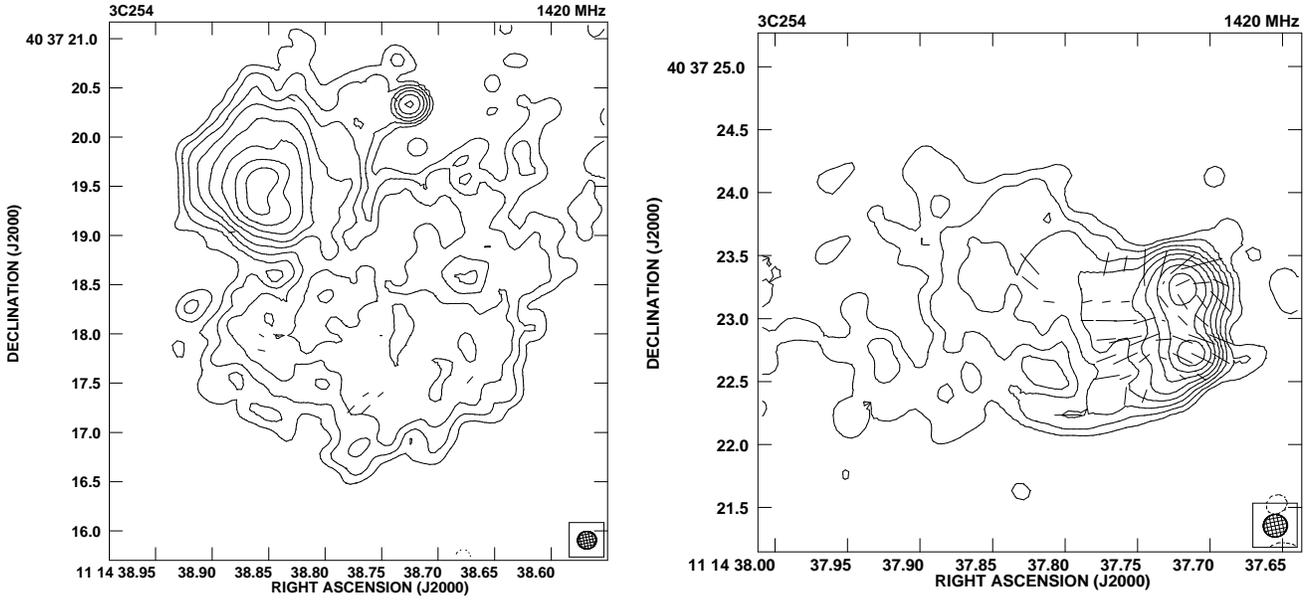

\begin{center}
\vbox{
\hspace{0.0cm}
         \hbox{
               \psfig{file=3C254-1420-ER.PS,width=3.3in,angle=0}
               \psfig{file=3C254-1420-WR.PS,width=3.6in,angle=-90}
              }
}
\end{center}
\caption{MERLIN images of the components of 3C254 at 1420 MHz with an angular resolution of
$\sim$0.18 arcsec.
The polarization E-vectors are superimposed on the total intensity contours.
Peak brightness: 164 and 143 mJy/beam for the western and eastern lobes respectively;
Contours: 0.7$\times$($-$1, 1, 2, 4, 8, 16 $\ldots$) mJy/beam. Polarization: 0.1 arcsec = 10 per cent.
}
\end{figure*}

\begin{figure*}
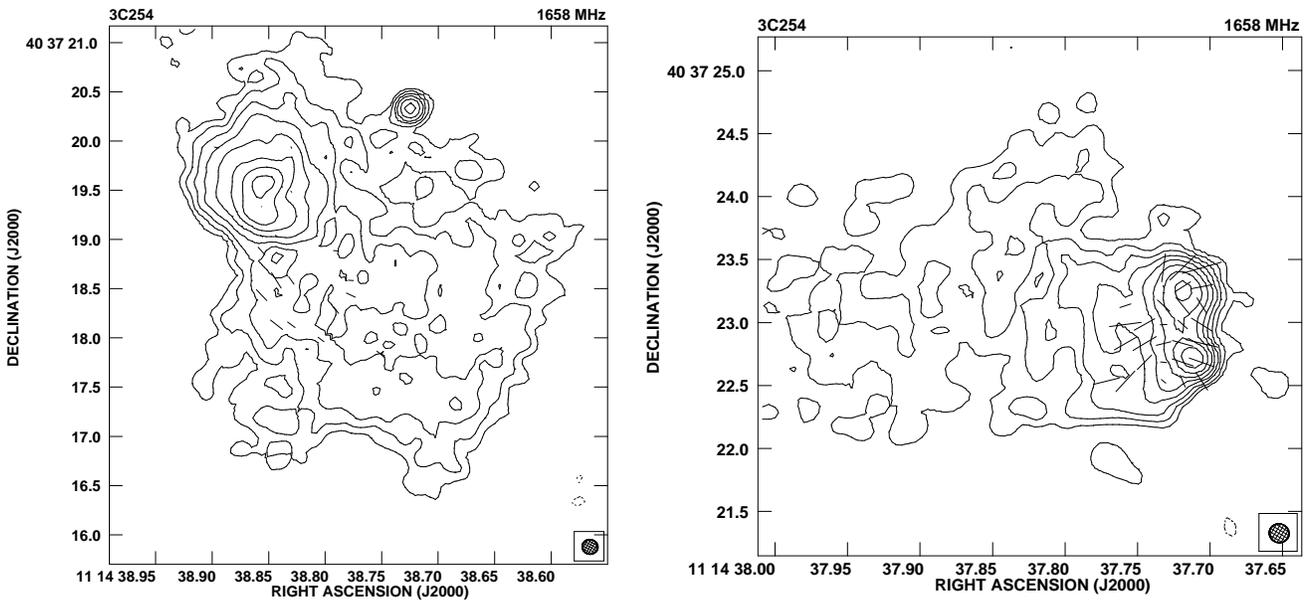

\begin{center}
\vbox{
\hspace{0.0cm}
         \hbox{
               \psfig{file=3C254-1658-ER.PS,width=3.3in,angle=0}
               \psfig{file=3C254-1658-WR.PS,width=3.6in,angle=-90}
              }
}
\end{center}
\caption{MERLIN images of the components of 3C254 at 1658 MHz with
an angular resolution of $\sim$0.15 arcsec.
The polarization E-vectors superimposed on the total intensity contours.
Peak brightness: 133 and 99 mJy/beam for the western and eastern components respectively;
Contours: 0.6$\times$($-$1, 1, 2, 4, 8, $\ldots$) mJy/beam. Polarization: 0.1 arcsec = 10 per cent.
}
\end{figure*}

\begin{table*}
\caption{Observational parameters and observed properties} 

\begin{tabular}{l r ccc c l lll rrl rr }
\hline
\multicolumn{6}{c}{Observational parameters} & \multicolumn{9}{c}{Observed properties} \\
Telescope & Freq. & \multicolumn{3}{c}{Resolution} & $\sigma$    &  Cmp   &  \multicolumn{3}{c}{RA(J2000)} & \multicolumn{3}{c}{Dec(J2000)} & S$_p$ & S$_t$  \\
   &    & maj. & min. & PA    &           &        &   &   &   &            &              &                    &       &           \\
  & MHz & $^{\prime\prime}$ & $^{\prime\prime}$ & $^\circ$ & mJy/b   &   & h & m & s & $^{\circ}$ &  $^{\prime}$ &  $^{\prime\prime}$ & mJy/b &  mJy       \\
(1)   &  (2)  &  \multicolumn{3}{c}{(3)}  & (4) & (5)  & \multicolumn{3}{c}{(6)} & \multicolumn{3}{c}{(7)}  &  (8) &  (9)    \\
\hline

MERLIN    & 1420  &  0.186 & 0.179 &  102  & 0.20      &  W       & 11 & 14 & 37.71   & 40 & 37 & 22.7  &    164 &  1104       \\
          &       &        &       &       &           &  C$^g$   &    &    & 38.73   &    &    & 20.3  &     26 &    28       \\
          &       &        &       &       &           &  E       &    &    & 38.86   &    &    & 19.5  &    143 &  1985       \\

MERLIN    & 1539  &  0.168 & 0.163 &   49  & 0.08      &  W       & 11 & 14 & 37.71   & 40 & 37 & 22.7  &    142 &   955       \\
          &       &        &       &       &           &  C$^g$   &    &    & 38.73   &    &    & 20.3  &     28 &    31       \\
          &       &        &       &       &           &  E       &    &    & 38.86   &    &    & 19.6  &    115 &  1747       \\

MERLIN    & 1658  &  0.155 & 0.149 &   60  & 0.19      &  W       & 11 & 14 & 37.71   & 40 & 37 & 22.7  &    133 &   959       \\
          &       &        &       &       &           &  C$^g$   &    &    & 38.72   &    &    & 20.3  &     26 &    29       \\
          &       &        &       &       &           &  E       &    &    & 38.86   &    &    & 19.6  &     99 &  1784       \\

MERLIN    & 4874  &  0.057 & 0.057 &       & 0.06      &  W       & 11 & 14 & 37.71   & 40 & 37 & 22.7  &     36 &   292       \\
          &       &        &       &       &           &  C$^g$   &    &    & 38.73   &    &    & 20.3  &     18 &    19       \\
          &       &        &       &       &           &  E       &    &    & 38.85   &    &    & 19.6  &    8.3 &   298       \\

MERLIN+   & 4874  &  0.100 & 0.100 &       & 0.05      &  W       & 11 & 14 & 37.71   & 40 & 37 & 22.7  &     51 &   330       \\
VLA       &       &        &       &       &           &  C$^g$   &    &    & 38.72   &    &    & 20.3  &     18 &    20       \\
          &       &        &       &       &           &  E       &    &    & 38.85   &    &    & 19.6  &     20 &   477 \\

\hline
\end{tabular}

Column 1: telescope used for the observations; column 2: observing frequency in MHz;
column 3: the angular resolution in arcsec and the position angle (PA) of the
restoring beam in deg;
column 4: the rms noise level in the image in units of mJy/beam; column 5: component
designation, with a superscript $g$ indicating that the values in columns 6 to 9 have
been estimated by fitting a two-dimensional Gaussian; columns 6 and 7: the right
ascension and declination of the peak of emission
of the component in J2000 co-ordinates; columns 8 and 9: the peak and total flux density
of the component in units of mJy/beam and mJy respectively. The errors in the
flux densities are approximately 10 per cent.

\end{table*}

\begin{figure*}
\begin{center}
\vbox{
\hspace{0.0cm}
\psfig{file=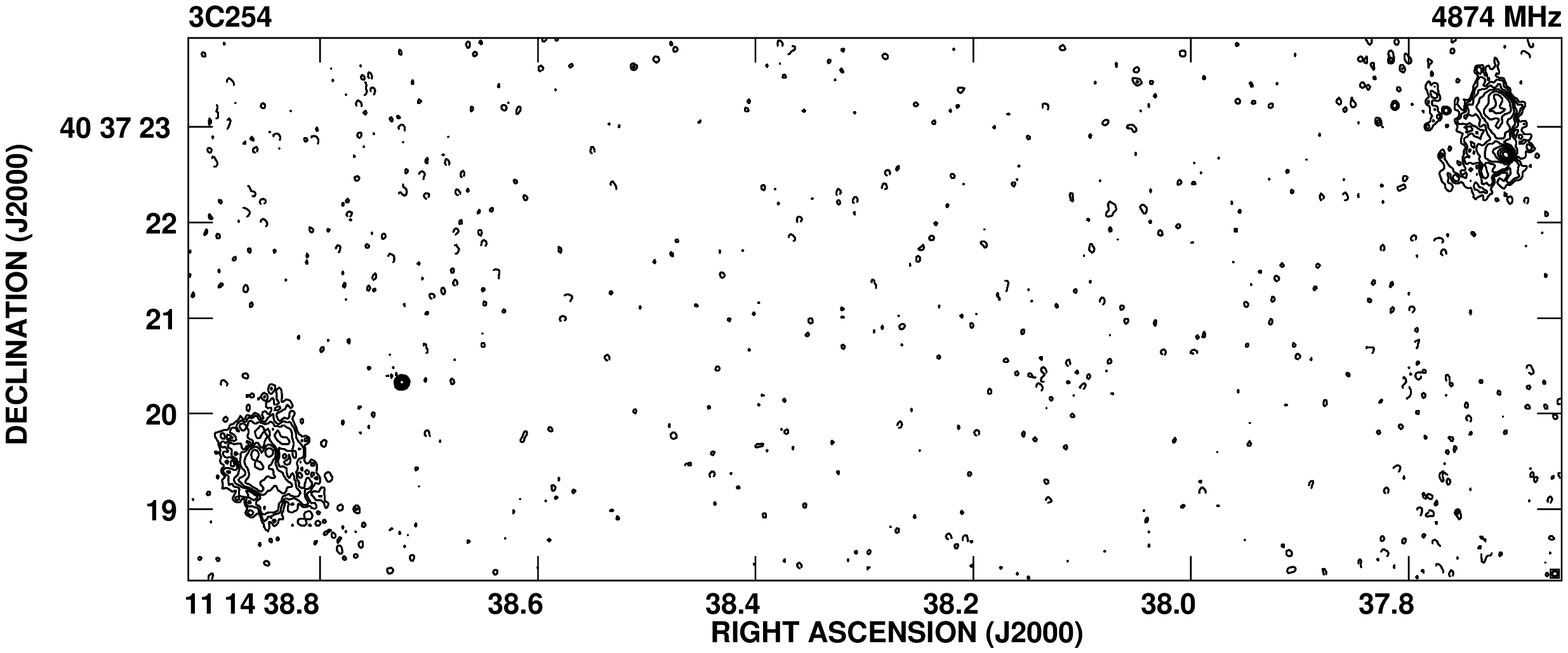,width=7.0in,angle=-90}
         \hbox{
               \psfig{file=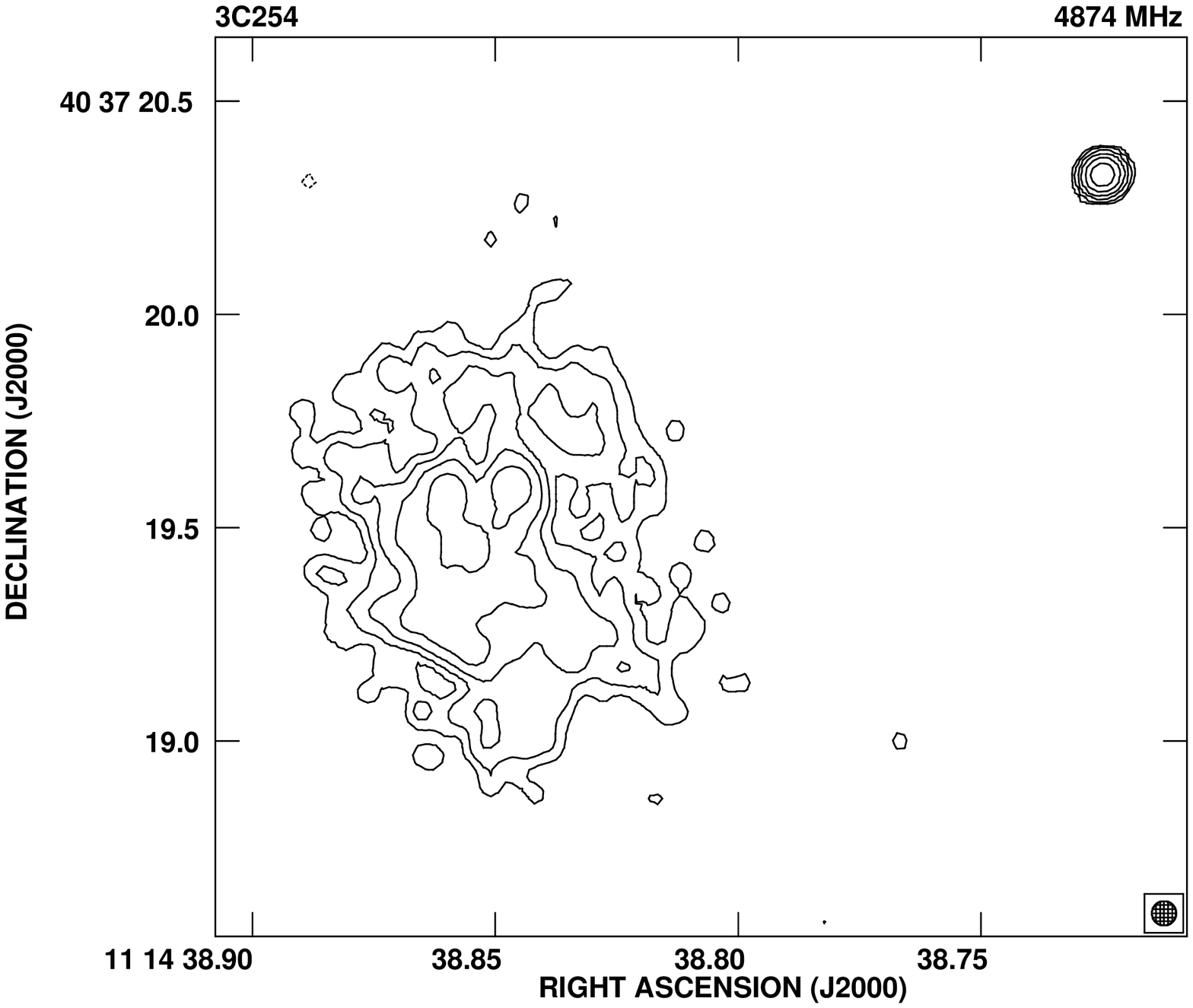,width=3.5in,angle=-90}
               \psfig{file=3C254-4874C-WR.PS,width=3.4in,angle=0}
              }
}
\end{center}
\caption{The MERLIN images of 3C254 at 4874 MHz with an angular resolution of 57 mas. The
total-intensity image is shown in the upper panel, while the lower panel shows the components
with the polarization E-vectors superimposed on the total intensity contours.
Peak brightness: 36 and 8.3 mJy/beam for the western and eastern components respectively; 
Contours: 0.17$\times$($-$1, 1, 2, 4, 8, $\ldots$) mJy/beam for the upper panel and 
0.3$\times$($-$1, 1, 2, 4, 8, $\ldots$) mJy/beam for the two lower panels. Polarization: 0.1 arcsec = 60 per cent.
}
\end{figure*}

\begin{figure*}
\begin{center}
\vbox{
\hspace{0.0cm}
\psfig{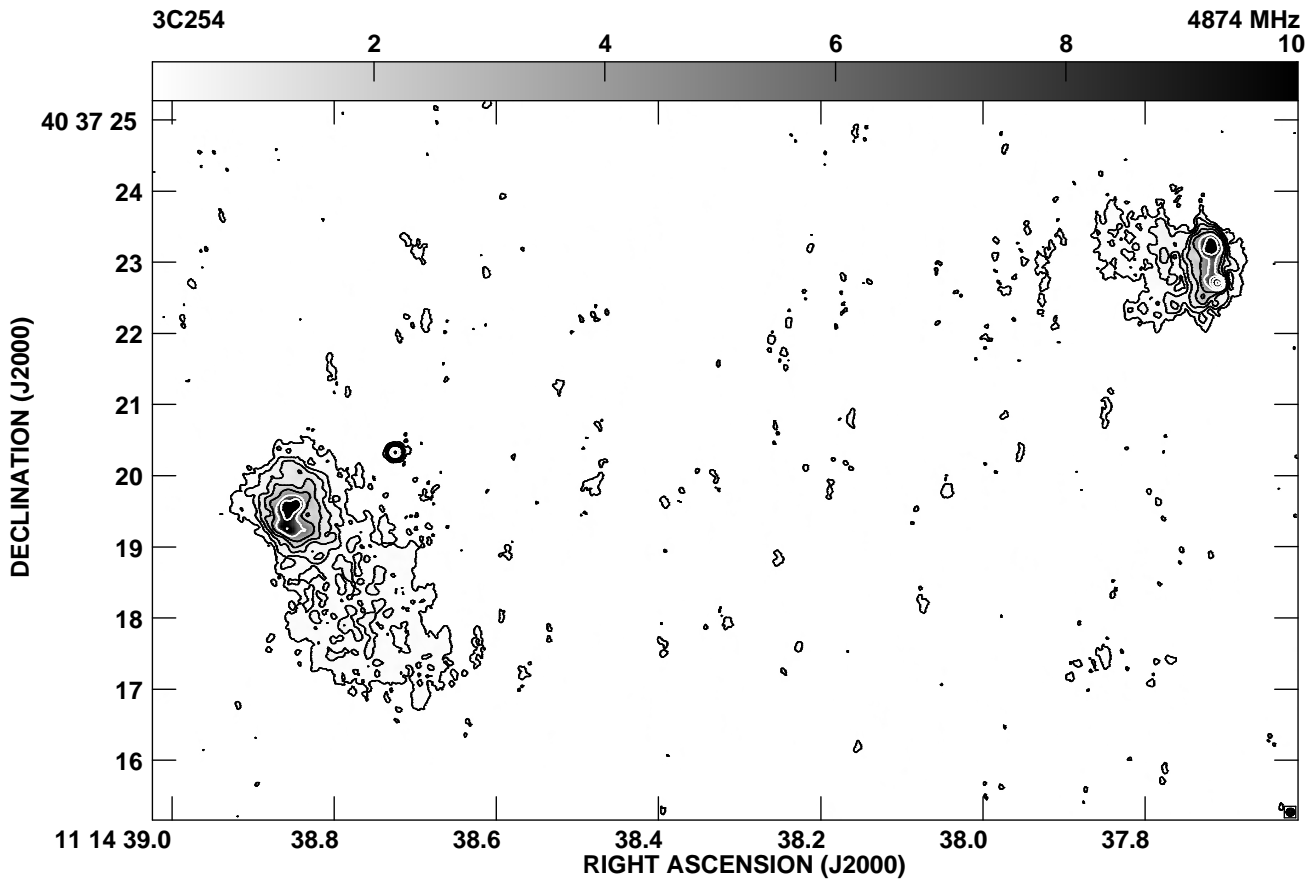}
     }
\end{center}
\caption{The MERLIN+VLA total-intensity image of 3C254 at 4874 MHz with an angular resolution of 100 mas. 
Peak brightness: 51 mJy/beam; Contours: 0.175$\times$($-$1, 1, 2, 4, 8, 
$\ldots$) mJy/beam. 
}
\end{figure*}

The MERLIN {\it C}-band image with an angular resolution of 57 mas (Fig. 4) shows the multiple
hotspot structure of the western lobe in more detail with the primary hotspot being
more compact with a peak brightness of 36 mJy/beam and peak polarization of
$\sim$30\%. The secondary hotspot is more diffuse with a peak brightness of $\sim$7 mJy/beam. 
For the small value of rotation measure, as discussed later in this Section, 
the inferred magnetic field lines at its northern edge appear to be parallel to the edge of the lobe.
The brightness and field structure of the primary and secondary hotspots are similar
to other known cases (e.g. Laing 1989; Hardcastle et al. 1997; Leahy et al. 1997), 
with the primary hotspot being identified with
the position at which the beam may be deflected by collision with a cloud of dense gas.
At this resolution the {\it C}-band image shows no bright hotspot on the
eastern lobe amongst several peaks of emission, the peak brightness being 8.3 mJy/beam.
There is no polarization detected from these features at 4874 MHz, an upper limit to
the degree of polarization being $\sim$2\%. 

The structures of the lobes on opposite sides of the core at the highest resolutions are
significantly different, with the western lobe being more typical of a hotspot in an FRII
source with a tail of emission while the eastern one is much more spherical and diffuse.
Hydrodynamic simulations of light, large-scale jets in a decreasing density profile, 
show that the jet bow shock undergoes two phases;
firstly a nearly spherical one and secondly the well-known cigar-shaped one
(cf. Krause 2002; Krause \& Camenzind 2002; Carvalho \& O'Dea 2002).  The shell-like structure 
of the eastern lobe is suggestive of 
the first phase of the development of the bow-shock. In this scenario, the eastern jet
has not yet entered the cigar phase and deposits its radio-emitting plasma in a large
part of the bubble, almost filling the region within the bounds of the bow shock. On 
the other hand, the western jet appears to be in the cigar phase and should therefore
have a fairly regular backflow around it, which flows back into the central parts diffusing
and mixing with the shocked external gas. 

Besides the asymmetry in the location and brightness of the outer
components, another striking feature of this source is the polarization asymmetry
(cf. Liu \& Pooley 1991). 
Since the images presented here with radio polarization information are at very 
different and high resolutions, it is not possible to derive reliable values of 
depolarization for the lobes. However, 
the depolarization of the western hotspot between {\it C}- and {\it L}-bands is
consistent with the measurements of Liu \& Pooley, 
while the eastern hotspot would appear to be strongly depolarized even at $\sim$5 GHz. The
rotation measure of the western hotspot between {\it C}- and {\it L}-bands is $\sim$$-$15
rad m$^{-2}$, which is consistent with values for other sources at similar Galactic
co-ordinates (e.g. Simard-Normandin, Kronberg \& Button 1981). 

\subsection{Which is the approaching side?}
One of the main objectives of these observations was to detect a radio jet in order to
identify the approaching component and to try to understand the extreme asymmetries observed
in this source. No radio jet has been detected to a brightness
limit of 0.08 mJy/beam pointing towards the western side in the best MERLIN {\it L}-band
image obtained from a combination of the two {\it L}-band frequencies.  It is difficult to put a similar
limit on a jet to the eastern side because of confusion with the lobe emission. In order to 
identify whether there is a jet on the eastern side, VLA {\it C}-band data
kindly provided by J. Riley and G. Pooley have been combined with the MERLIN data to achieve the best
sensitivity with an adequate resolution. The combined MERLIN+VLA image (Fig. 5), which has an
angular resolution of 0.1 arcsec and an rms noise of 0.05 mJy/beam, shows no evidence of any 
jet on either the western or eastern side.

Therefore, from jet sidedness it is difficult to identify the approaching and receding
sides of the source. The blue-shifted optical emission lines, which have been interpreted
as being caused by a collision of the jet with a dense cloud,  have led  Bremer (1997)
to suggest that the eastern lobe is approaching us. This is a plausible scenario and,
assuming this to be true,
3C254 is inconsistent with the Laing-Garrington effect (Laing 1988; Garrington et al. 
1988). This is not surprising if the external environment is very asymmetric with the
eastern jet interacting with dense gas which slows down the jet and also
depolarizes the radio emission. The strong depolarisation in the eastern lobe may also
be partly due to the lobe being within the associated host galaxy since its separation from
the core is only $\sim$12 kpc while the western lobe is in a more tenuous medium.

\subsection{Are the oppositely-directed jets symmetric?} 
It might be expected that the brightness of the eastern hotspot would be greater than that of the western hotspot because it is interacting with a much denser
medium, which should lead to a greater dissipation of energy and higher efficiency of conversion
of beam energy into radio emission (e.g. Eilek \& Shore 1989; Gopal-Krishna \& Wiita
1991; Jeyakumar et al. 2005). Additionally, if it is indeed on the approaching side, its brightness may be enhanced by relativistic beaming (Rees 1966). However, the ratio of the  
peak brightness of the hotspots on the eastern side to that on the western side at the highest 
resolution of 57 mas is 0.23. This is just the reverse of what might normally
be expected, suggesting that the oppositely-directed beams may be intrinsically asymmetric.
Evidence of intrinsically asymmetric jets have come from studies of individual sources
such as B0500+630 which has a complete absence of hotspots on one side (Saikia et al. 1996).
Other examples of such sources have been given by  Gopal-Krishna \& Wiita (2000). 

\section{Concluding remarks}
The radio galaxy, 3C254, is one of the most asymmetric radio sources in terms of 
the location of the outer lobes of radio emission relative to the nucleus of the host
galaxy. MERLIN observations of the source have been presented at {\it L}- and {\it C}-bands with
resolutions ranging from 0.18 (1.3 kpc) to 0.057 arcsec (0.4 kpc) respectively. These sensitive,
high-resolution observations do not show any evidence of a jet on either the western
or eastern side of the source, making it difficult to identify the approaching side
on the basis of jet sidedness. Identifying the eastern side as the approaching one
on the basis of the blue-shifted emission lines being pushed outwards by interaction 
with the radio jet (Bremer 1997), it appears that while the large-scale asymmetries may be caused by
interaction with an asymmetric external environment, the small-scale asymmetries in 
the hotspots are probably the result of an intrinsic asymmetry in the jets on opposite sides.
The usual assumption of oppositely-directed symmetric jets does not appear to be
universally applicable, but it may be possible to establish this in only the most
asymmetric of sources.

\section*{Acknowledgments}
We thank Julia Riley and Guy Pooley for giving us the calibrated VLA A-array data at 5 GHz, 
and an anonymous referee, Graham Smith and Alan Pedlar for several helpful comments.
One of us (DJS) would like to thank the PPARC Visitors Programme at Jodrell Bank Observatory,
Andrew Lyne, Director, for use of the facilities at the Observatory, and Peter Thomasson for 
hospitality while this
work was done.  MERLIN is a U.K. National Facility operated by the University of Manchester 
on behalf of PPARC.  The Very Large Array is operated by the National Radio Astronomy
Observatory for Associated Universities Inc. under a co-operative
agreement with the National Science Foundation.  
This research has made use of the NASA/IPAC extragalactic database (NED)
which is operated by the Jet Propulsion Laboratory, Caltech, under contract
with the National Aeronautics and Space Administration.

{}


\begin{thebibliography}{}

\bibitem[]{} Baars  J.W.M., Genzel  R., Pauliny-Toth I.I.K., Witzel A., 1977, A\&A, 61, 99
\bibitem[]{} Barthel P.D., 1989, ApJ, 336, 606
\bibitem[]{} Bremer M.N., 1997, MNRAS, 284, 126
\bibitem[]{} Carvalho, J.C.,  O'Dea C.P., 2002, ApJS, 141, 371 
\bibitem[]{} Crawford C.S., Vanderriest C., 1997, MNRAS, 285, 580
\bibitem[]{} Donahue M., Daly R.A., Horner D.J., 2003, ApJ, 584, 643
\bibitem[]{} Eilek J.A., Shore S.N., 1989, ApJ, 342, 187
\bibitem[]{} Forbes D.A., Fabian A.C., Johnstone R.M., Crawford C.S, 1990, MNRAS, 244, 680
\bibitem[]{} Garrington S.T., Leahy J.P., Conway R.G., Laing R.A., 1988, Nature, 331, 147
\bibitem[]{} Gopal-Krishna, Wiita P.J., 1991, ApJ, 373, 325
\bibitem[]{} Gopal-Krishna, Wiita P.J., 2000, A\&A, 363, 507
\bibitem[]{} Hardcastle M.J., Alexander P., Pooley G.G., Riley J.M., 1997, MNRAS, 288, 859
\bibitem[]{} Ishwara-Chandra C.H., Saikia D.J., McCarthy P.J., van Breugel W.J.M., 2001, MNRAS, 323, 460
\bibitem[]{} Jeyakumar S., Wiita P.J., Saikia D.J., Hooda J.S., 2005, A\&A, 432, 823
\bibitem[]{} Krause M., 2002, A\&A, 386, L1 
\bibitem[]{} Krause M., Camenzind M., 2002, In Active Galactic Nuclei: from Central Engine to Host 
             Galaxy, eds.  S. Collin, F. Combes and I. Shlosman, ASP Conf. Sr., in press 
\bibitem[]{} Laing R.A., 1988, Nature, 331, 149
\bibitem[]{} Laing R.A., 1989, in  Hot spots in extragalactic radio sources, eds K. Meisenheimer, H.-J. Roeser, Springer-Verlag, 
             Berlin, p. 27.
\bibitem[]{} Leahy J.P., Black A.R.S., Dennett-Thorpe J., Hardcastle M.J., Komissarov S., Perley R.A., 
             Riley J.M., Scheuer P.A.G., 1997, MNRAS, 291, 20
\bibitem[]{} Liu R., Pooley G., 1991, MNRAS, 249, 343
\bibitem[]{} Liu R., Pooley G., Riley J.M., 1992, MNRAS, 257, 545
\bibitem[]{} McCarthy P.J., van Breugel W., Kapahi V.K., 1991, ApJ, 371, 478
\bibitem[]{} Owen F.N., Puschell J.J., 1984, AJ, 89, 932
\bibitem[]{} Patnaik A.R., Browne I.W.A., Wilkinson P.N.,  Wrobel J.M., 1992,  MNRAS, 254, 655
\bibitem[]{} Pedelty J.A., Rudnick L., McCarthy P.J., Spinrad H., 1989, AJ, 97, 647
\bibitem[]{} Rees M.J., 1966, Nature, 211, 468
\bibitem[]{} Reid A., Shone D.L., Akujor C.E., Browne I.W.A., Murphy D.W., Pedelty J., Rudnick L., Walsh D., 
             1995, A\&AS, 110, 213
\bibitem[]{} Saikia D.J., Junor W., Cornwell T.J., Muxlow T.W.B., Shastri P., 1990, MNRAS, 245, 408
\bibitem[]{} Saikia D.J., Thomasson P., Jackson N., Salter C.J., Junor W., 1996, MNRAS, 282, 837
\bibitem[]{} Simard-Normandin M., Kronberg P.P., Button S., 1981, ApJS, 45, 97
\bibitem[]{} Spergel D.N., et al. 2003, ApJS, 148, 175 
\bibitem[]{} Thomasson P., Saikia D.J., Muxlow T.W.B., 2003, MNRAS, 341, 91 

\end{thebibliography}
\end{document}